\documentclass[a4paper,superscriptaddress,twocolumn,prb,aps,showpacs,floatfix]{revtex4}

\usepackage{epsfig,url}
\usepackage{epsf}
\usepackage{graphicx}
\usepackage{amsmath,amssymb}
\usepackage{booktabs}  % nice tables
\usepackage{dcolumn}
\usepackage{rotating}
\usepackage{bm}        % for bold math
\usepackage{bbm}       % for nice three dimensional spaces
\begin{document}

\title{The planar-to-tubular structural transition in boron clusters from
  optical absorption}

\author{Miguel A.\,L. Marques}
\affiliation{Institut de Min\'eralogie et de Physique des Milieux Condens\'es,
Universit\'e Pierre et Marie Curie - Paris VI, 140 rue de Lourmel,
75015 Paris, France}

\author{Silvana Botti}
\affiliation{Laboratoire des Solides Irradi\'es, CNRS-CEA-\'Ecole Polytechnique,
91128 Palaiseau, France}

\date{\today}

\begin{abstract}
  The optical response of the lowest energy isomers of the B$_{20}$
  family is calculated using time-dependent density functional theory
  within a real-space, real-time scheme. Significant differences are
  found among the absorption spectra of the clusters studied. We show
  that these differences can be easily related to changes in the
  overall geometry.  Optical spectroscopy is thus an efficient tool to
  characterize the planar to tubular structural transition, known to
  be present in these boron based systems.
\end{abstract}


\maketitle

For the past years, boron nanostructures have attracted the attention of both
theoretical and experimental physicists. This is due to the remarkable
properties of boron, that make it an unique element in the periodic table, with
important technological applications.\cite{cotton99,greenwood97} Boron is characterized by a short covalent
radius and has the tendency to form strong and directional chemical
bonds.\cite{cotton99,greenwood97} These characteristics lead to a large
diversity of boron nanostructures --
clusters,\cite{hanley88,laplaca92,zhai02,zhai03,kiran05} nanowires,\cite{meng03}
and nanotubes\cite{boustani99,ciuparu04} -- that have already been observed.

Many experimental studies of small boron-based clusters have been
performed in the last decade, namely using mass and photo-electron
spectroscopies.\cite{hanley88,laplaca92,zhai02,kiran05} However, we
still have very limited information regarding the geometries and
electronic properties of these systems. From the theoretical point of
view, there have been extensive {\it ab initio} quantum chemical and
density functional calculations about the structural properties of
neutral,\cite{hanley88,kato91-kato92-kato93,ray92,tang93,boustani97,niu97,boustani99bis-chacko03,zhai02,kiran05}
cationic,\cite{boustani94,ricca96,niu97} and anionic
clusters.\cite{zhai02,kiran05}  The findings were fairly surprising.
In fact, bulk boron appears in several crystalline and amorphous
phases, the best know of which are the $\alpha$- and
$\beta$-rombohedral, and the $\alpha$-tetragonal, also known as low
temperature or red boron. In these three phases, boron is arranged in
B$_{12}$ icosahedra\cite{cotton99,perkins96,hubert98} (in the
$\alpha$-tetragonal phase these icosahedra are slightly distorted).
However, the small clusters appear in four distinct
shapes:\cite{boustani99bis-chacko03} convex, spherical, quasi-planar,
and nanotubular -- totally unrelated to the B$_{12}$ icosahedra.

The most stable members of the B$_n$ family with $n\lesssim 20$ are
known to be
planar.\cite{hanley88,kato91-kato92-kato93,ray92,tang93,boustani97,niu97,zhai02}
Recent calculations showed that B$_n$ clusters with $n=24$ and $n=32$
prefer tubular structures.\cite{boustani99bis-chacko03} In a recent
study\cite{kiran05} Kiran {\it et al.} placed the transition between
these two topologies at $n=20$. However, their results were not
totally conclusive: while the theoretical calculations (using density
functional theory at the B3LYP/6-311+G$^*$ level) yielded a double
ring arrangement as the lowest energy isomer of B$_{20}$ and
B$^{-}_{20}$, the experimental photo-electron spectra of anionic
aggregates of the same size were only compatible with planar
structures. Such a incongruity can be explained by the difficulties
associated both to the experimental and numerical
techniques. Experimentally, the clusters were produced by laser
evaporation of a disk target, with the formation of the fragments
being mainly controlled by kinematics. The situation is similar, e.g.,
to the case of C$_{20}$: while the rings are easily
produced by laser evaporation, the apparently more stable bowl and cage
arrangements do not form spontaneously, but can only be obtained by
fairly sophisticated chemical techniques.\cite{prinzbach00} On the
other hand, the theoretical determination of the lowest energy isomer
is a very arduous task to current computational material science. The
reason is twofold: i)~with increasing number of atoms the number of
metastable isomers increases exponentially; ii)~and often the energy
difference between competing structures becomes quite small. In the
case of B$^-_{20}$, e.g., the four lowest lying isomers are separated
by less than 0.3\,eV (15\,meV per atom).\cite{kiran05} Clearly, this
precision is beyond current density functional and quantum chemical
methods for systems of this size.  Once again, C$_{20}$ is an
illustrative example: to unveil its elusive ground state, scientists
have tried density functional theory at various
levels,\cite{grossman95,jones97} quantum Monte Carlo,\cite{grossman95}
coupled cluster,\cite{bylaska96} etc., without reaching any agreement
regarding the energy ordering of the isomers.

In view of this situation, instead of relying on a single number,
i.e., the total energy, in order to characterize a system, it is
better to resort to the several spectroscopic tools available to both
computational and experimental physics. It was, for example, the case
of C$_{20}$, where the bowl and the cage isomers were identified by
comparing the vibrionic fine structures in photo-electron spectra with
numerical calculations.\cite{saito01} In this paper we propose the use
of another spectroscopic tool, namely optical absorption, to provide
clear signatures of the distinct B$_{20}$ isomers. In fact,
when the orbitals involved in the optical transitions are quite
extended, as in the case of these boron clusters, optical spectroscopy
in the visible and near-ultraviolet (near-UV) range turns out to be a
rather sensitive probe of the overall shape of the system.
In contrast to total energy differences, the absorption spectra are usually quite insensitive
to small changes in the geometry, to the parametrization
used for the exchange-correlation potential, and to changes in the 
pseudopotentials. This makes this tool valuable in studying structural transitions.

\begin{figure}[t]
\begin{center}
\epsfig{file=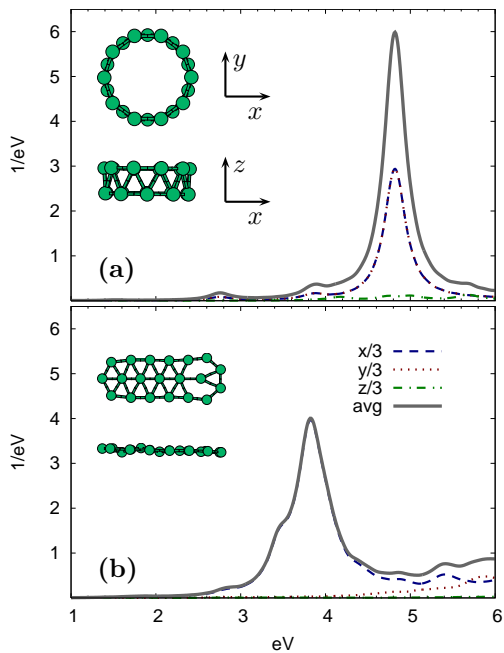,scale=1.0,angle=0.0}
\caption{\label{Fig:spectra1}
  (Color online) Dipole strength function for the different isomers of
  B$_{20}$. The corresponding geometries are shown in each panel. The gray
  (solid) line represents the absorption averaged over the three polarization
  directions; the blue (dashed), the red (dotted), and the green (dashed-dotted)
  lines indicate, respectively, the absorption in the $x$, $y$, and $z$
  directions. (These three last curves have been divided by 3.) The reference
  frame used is depicted in the top panel.}
\end{center}
\end{figure}

\begin{figure}[t]
\begin{center}
\epsfig{file=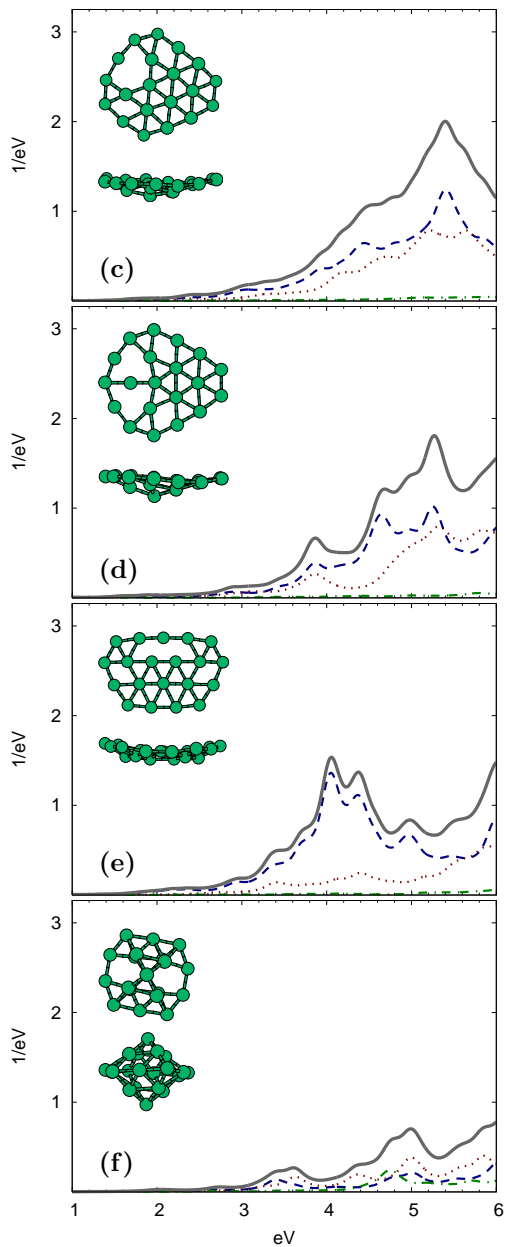,scale=1.0,angle=0.0}
\caption{\label{Fig:spectra2}
  (Color online) The same as Fig.~\ref{Fig:spectra1} for the rest of the
  members of the B$_{20}$ family studied in this work. Note the different vertical scale with
  respect to Fig.~\ref{Fig:spectra1}.  }
\end{center}
\end{figure}

Our procedure relies on density functional theory and is
briefly explained in the following. The geometries were optimized with the
computer code siesta,\cite{siesta} employing a double $\zeta$ with polarization
basis set. From the optimized geometries, we then obtain the optical spectra within
time-dependent density functional theory (TDDFT) in real time, as implemented in
the computer code {\tt octopus},\cite{octopus} using the PBE parametrization\cite{pbe}
in the adiabatic approximation for the exchange-correlation potential.
For a technical description of this method we refer to Refs.~\onlinecite{octopus} and
\onlinecite{castro04}. This approach has already been used for the study of metal and
semiconducting clusters,\cite{marques01,castro04} aromatic
molecules,\cite{yabana99} protein chromophores,\cite{marques03}
etc. Moreover, it proved quite successful in distinguishing the
different isomers of C$_{20}$.\cite{castro02} When experimental
spectra were available for a direct comparison, TDDFT within an
adiabatic local density (LDA) or an adiabatic generalized gradient (GGA)
approximation to the exchange-correlation functional reproduced the
low energy peaks of the optical spectra with an accuracy of
0.1\,eV.\cite{castro04} This is in contrast with taking the differences
of the eigenvalues of the HOMO and LUMO orbitals, which would give 
peaks at lower frequencies in disagreement with the experimental spectra.\cite{castro02} 

To represent the wave-functions in real space we used a uniform
grid with a spacing of 0.23\,\AA, and a box extending 5\,\AA\ from the outer nuclear
positions. A time step of 0.0013\,fs assures the stability of the time
propagation, and a total propagation time of 24.5\,fs allows a resolution of
about 0.1\,eV in the resulting spectrum.

The isomers of B$_{20}$ considered in this work can be divided in three
broad categories (see Figs.~\ref{Fig:spectra1} and
\ref{Fig:spectra2}): {\bf a} -- a nanotubular structure; {\bf b} to {\bf
e} -- quasi-two dimensional (2D) clusters; and {\bf f} -- a three
dimensional (3D) cage, that we obtained through a minimization
procedure starting from a dodecahedron (a B$_{20}$ fullerene-like
structure). The geometries that we selected include the lowest energy
isomers found in Ref.~\onlinecite{kiran05}. Furthermore, we decided to
add the cage isomer {\bf f}, in order to reach more general
conclusions on the effects of the 2D-to-3D transition on the optical
response.

Total energies calculated at the relaxed geometries both with siesta\cite{siesta}
and {\tt octopus}\cite{octopus} confirm that the tubular structure {\bf a} is the
lowest lying isomer of the B$_{20}$. This is in agreement with
the previous findings of Ref.~\onlinecite{kiran05}, obtained using a
B3LYP functional and a Gaussian basis set. In our calculation then
follow isomers {\bf c} ($\Delta E$=0.43\,eV), {\bf d} ($\Delta
E$=0.54\,eV), {\bf f} ($\Delta E$=0.62\,eV), {\bf b} ($\Delta
E$=0.81\,eV), and {\bf e} ($\Delta E$=0.91\,eV).  Note that the
ordering is consistent with the one proposed in
Ref.~\onlinecite{kiran05}, except for cluster {\bf f}, which was not
considered in Ref.~\onlinecite{kiran05} as a low energy candidate for
B$_{20}$. This discordance can be attributed to the basis
set used in Ref.~\onlinecite{kiran05}, or to the changes in the 
exchange-correlation functionals or pseudopotentials. 
Once again this reveals the danger of relying solely on the 
total energy to characterize such structural transitions.

The results of our calculations are summarized in Figs.~\ref{Fig:spectra1} and \ref{Fig:spectra2},
where we plot the computed dipole strength for six low-lying members of the
B$_{20}$ family. Besides the absorption spectra averaged over the three polarization directions
(solid gray lines), we also depict the three components for light polarized along the Cartesian axis (see
Fig.~\ref{Fig:spectra1}\,{\bf a} for the definition of the axis).

From the observation of Figs.~\ref{Fig:spectra1} and
\ref{Fig:spectra2}, it is clear that the isomers of B$_{20}$ have
strikingly different absorption properties in the visible and near-UV,
the range most easily accessible by experimental techniques. The
tubular cluster {\bf a} exhibits a large sharp peak at 4.8\,eV, and
two small peaks at around 2.7 and 3.9\,eV.  
Also the quasi-planar isomer {\bf b} absorbs light at well
defined frequencies, with a main peak at 3.8\,eV and a shoulder at
3.5\,eV. The other quasi-2D structures {\bf c}, {\bf d}, and {\bf e}
no longer exhibit sharp peaks, but instead present broad
features. This is easily explained by noting that these clusters have
a larger distribution of bond lengths, due to the lack of symmetry
around the boron centers. This is particularly evident in isomer {\bf
c}, resulting in an almost structureless spectrum that grows
monotonically until around 5.4\,eV.  Nevertheless, the differences
between these three spectra are still sufficient to distinguish among
them: cluster~{\bf d} has a well defined peak at 3.9\,eV, and the
double peak at 4.0 and 4.4\,eV is a clear signature of cluster~{\bf
e}. Isomer {\bf f} absorbs weakly visible and near-UV light. All
quasi-2D isomers do not show any appreciable response for light
polarization perpendicular to their plane;
moreover, the response is
stronger along the direction in which the cluster is more extended.
Also in the case in which the plane is rolled up to form a tube,
the contributions to the peaks come almost exclusively from absorption
of light polarized perpendicularly to the axis of the tube.

\begin{figure}[t]
\begin{center}
\begin{tabular}{cc}
\epsfig{file=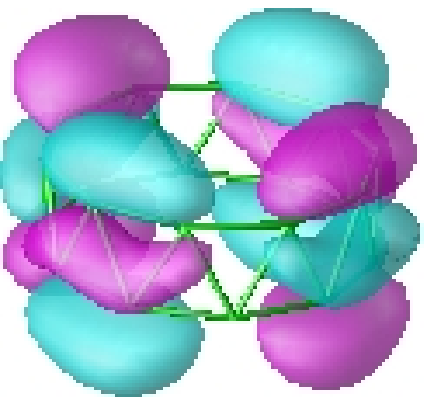,scale=0.5,angle=0.0} &
\epsfig{file=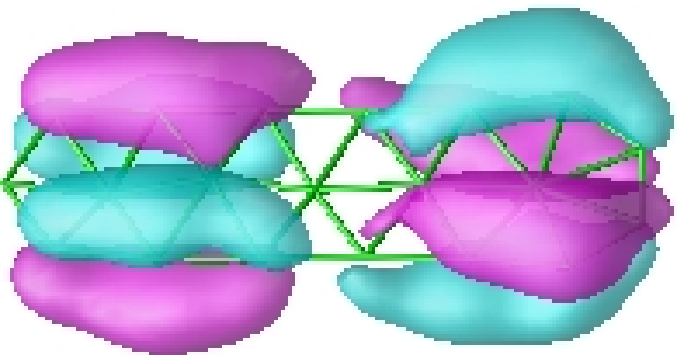,scale=0.5,angle=0.0} \\
\epsfig{file=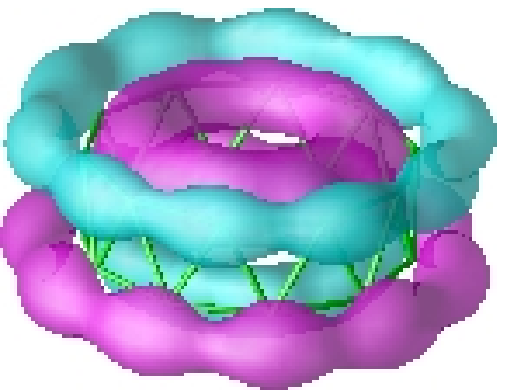,scale=0.5,angle=0.0} &
\epsfig{file=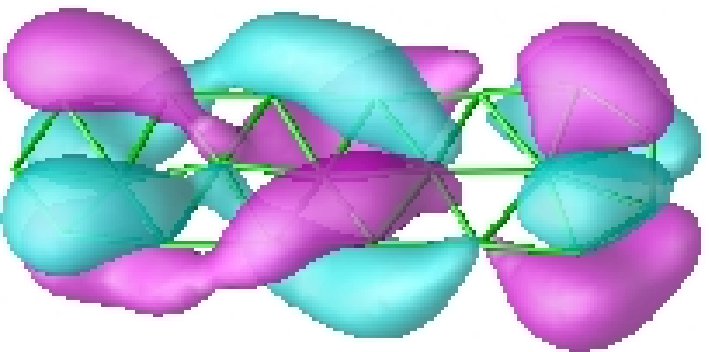,scale=0.5,angle=0.0} \\
\end{tabular}
\caption{\label{Fig:states}
  (Color online) HOMO (top) and LUMO (bottom) states of the isomers {\bf a} (left) and {\bf b}
  (right). The magenta (cyan) isosurface corresponds to the positive (negative)
  part of the wave-functions.  }
\end{center}
\end{figure}

In order to understand better the absorption in the planar and tubular
isomers, we can learn from the shape of the Kohn-Sham orbitals (see
Fig.~\ref{Fig:states}). Both the HOMO and the LUMO, and several other
states relevant to absorption, have predominantly $\pi$ symmetry (with
some $\sigma$ admixture due to the breaking of planarity, which is
particularly important in the case of the tubular shape). In this kind
of compounds, the optical gap decreases with increasing length of the
$\pi$ system, and consequent increase of delocalization of the valence
electrons. This is exactly what we observe in our spectra of
planarlike isomers: isomer~{\bf b}, extending to almost 10\,\AA\ in
the $x$ direction absorbs at the lowest energy, followed by
isomer~{\bf e}, and finally {\bf c} and {\bf d}. As already mentioned,
these isomers are almost transparent for light polarized perpendicular
to the plane ($z$ direction in the figure). In fact, the light
polarized in the $z$ direction cannot excite the $\pi$--$\pi^*$
transitions and thus gives an almost negligible contribution to the
spectra.  In isomer~{\bf a}, there is a stronger contribution of the
$\sigma$ bonds to the relevant orbitals due to the
curvature of the tube wall (see Fig.~\ref{Fig:states}), which induces
a considerable blue-shift of the main peak. Finally, we 
discuss the 3D cage {\bf f} (see Fig.~\ref{Fig:spectra2}). As its
bonds have a predominantly $\sigma$ character, it does not exhibit
strong absorption below 6\,eV. As in the case of the irregular
quasi-2D clusters, the large number of inequivalent boron atoms
is responsible for the absence of well defined peaks.

The present results for B$_{20}$, together with the analogous results
for C$_{20}$ isomers,\cite{castro02} lead to some conclusions of wider validity. 
Two factors contribute to the position of the peaks and 
the overall shape of the spectra: i)~the extension of the
$\pi$ system, which is directly related to the dimensionality and
basic geometry of the cluster, determines the frequencies of strong
absorption; and ii)~the distribution of bond-lengths, related to the
number of inequivalent boron atoms, determines if the spectrum is
composed of sharp peaks or broad features. These factors are quite
general, and are thus expected to apply to boron clusters of different
sizes and any other class of clusters with similar chemical bonds.

In conclusion, we have shown how optical spectroscopy can be used to
distinguish without ambiguity between the different low-energy members
of B$_{20}$ family. The spectra in the visible and near-UV are very
sensitive to the overall shape of the isomers. This is a general
property that can be easily explained by simple geometrical arguments.
In particular, the most stable neutral B$_{20}$ isomer, i.e. the
tubular cluster, can be unequivocally identified due to the presence
of a very sharp resonance at about 4.8\,eV.  For this reason, we
believe that optical spectroscopy can be an extremely efficient tool
to study structural transitions in clusters. In the case of B$_{n}$
aggregates, the comparison between experimental and computed
absorption spectra, could bring a definitive answer to the unsolved
question of the critical size at which the transition between the
planar and tubular structure occurs.

The authors would like to thank A. Castro, X. L\'{o}pez, L. Reining, A. Rubio,
and A. Seitsonen for useful suggestions and comments.  MALM was supported by the
Marie Curie Actions of the European Commission, Contract No. MEIF-CT-2004-010384.
The authors also acknowledge partial support by the EC Network of Excellence
NANOQUANTA (NMP4-CT-2004-500198). All computations were performed in the
Laborat\'orio de Computa\c{c}\~ao Avan\c{c}ada of the University of Coimbra
(Portugal).

%%%%%%%%%%%%%%%%%%%%

\end{document}